\documentclass{jpsj3}
\usepackage{amsmath}
\usepackage{graphicx}
\usepackage{amssymb}

\usepackage{gt21jpsj}
\usepackage{color} 
\usepackage{graphicx}
\usepackage{bm}
\usepackage{amsmath}

\def\newblock{\hskip .11em plus .33em minus .07em} 
\begin{document}
\title{
Nonlocality of electrically-induced spin accumulation in chiral metals
}
\author{Gen Tatara} 
\inst
{RIKEN Center for Emergent Matter Science (CEMS)
and RIKEN Cluster for Pioneering Research (CPR), 
2-1 Hirosawa, Wako, Saitama, 351-0198 Japan}
\date{\today}
\abst{
Spin accumulation induced by an electric field in chiral electron system is  investigated based on a linear response theory. It is shown that the spin response function has a spatially uniform component due to the chiral angular momentum generation effect, resulting in a nonlocal spin generation.
The decay length of spin profile is proportional to the electron elastic mean free path.
}
\maketitle

\newcommand{\Qv}{{\bm Q}}

Efficient transport of electron spin in solids is essential for spintronics.
For a long-range transport, insulating materials like YIG and NiO$_2$  have been known to be good spin conductors due to its weak damping \cite{Jungfleisch18,Lebrun20}.
In most spin transport experiments, spin current generation and detection are carried out electrically by use of the direct and inverse spin Hall effects \cite{Hirsch99,Saitoh06}.
Theoretically, spin Hall effect has been argued mostly in the context of spin Hall conductivity  described by the uniform component of the response function of spin current to an applied electric current.
The description has a fundamental ambiguity in the definition of spin current as has been widely argued \cite{Shi06,Shitade22}.

The physical observable of spin Hall effect is the spin accumulation, and theoretical formulation in terms of induced spin instead of spin current \cite{DyakonovPhysLett71}is free from such ambiguity.
To describe spin accumulation based on the linear response theory, the spin response function needs to be studied beyond the uniform ($q=0$) approximation, as the spin Hall-induced spin accumulation is non uniform.
It was demonstrated in Ref. \cite{TataraSH18} that the response function for the random impurity spin-orbit interaction expanded to the linear order of wave vector $q$ correctly reproduces the spin spatial profile of the conventional theory of spin Hall effect.
In the case of Rashba spin-orbit interaction, the response was shown to have a uniform component corresponding to the Rashba-Edelstein effect.
The structure of spin Hall response function in the $q$ space was later studied and demonstrating that the spin Hall response function has a broad weight at low $q$ \cite{Tatara22}.

Similarly, the inverse spin Hall effect is equivalent to an electric detection of a local spin accumulation \cite{TataraSH18}, although it is conventionally interpreted as an electric detection of a spin current flowing into a heavy metal lead.
Spin transport measurements by use of the direct and inverse spin Hall effects like in  Ref. \cite{Lebrun18} therefore correspond to  an electric measurement of the magnetic susceptibility \cite{TataraSH18,TataraAF19}.
In this context, nonlocal ``transport'' of spin is natural, because the susceptibility as function of an external wave vector ($q$)  generally has a weight near $q=0$.
The decay length of spin accumulation, determined by the width of the response function near $q=0$ \cite{TataraAF19}, is governed by the spin relaxation effect and is proportional to the elastic electron mean free path in the case of Elliott-Yafet relaxation \cite{Elliott54,Yafet83,Boross13} caused by the spin-orbit interaction.

Of particular recent interest in the context of spin transport are  chiral systems, where left- and right-handedness are distinguishable due to broken spatial inversion symmetry.
A strict definition of chiral requires in addition the invariance under time-reversal \cite{Barron86}.
Concerning conduction electrons, a typical chiral interaction is
a spin-orbit interaction linear in the electron wave vector $\kv$ and spin $\sigmav$, $\kv\cdot\sigmav$, called the Weyl spin-orbit interaction.
In fact, it was demonstrated that tight-binding model for electrons in a chiral crystal structure is mapped to a Weyl model, and the orbital Edelstein effect was discussed based on the model \cite{Yoda18}.
The Weyl electron system was recently pointed out theoretically to have a bulk orbital angular momentum and spin response to an applied electric field \cite{Toshio20,Funaki21}.
The results suggest that electrically-induced spin accumulation in chiral systems has a nonlocal nature.
Here, we explore the spin response in chiral system and demonstrate that the electrically-induced spin accumulation is a nonlocal effect due to the chiral nature.

Recent experiments revealed that chiral conductors such as CrNb$_3$S$_6$ \cite{Inui20}, NbSi$_2$ and TaSi$_2$ \cite{Shiota21} show large spin polarization under an applied electric current, the effect called the chirality-induced spin selectivity (CISS). Besides a large spin polarization, long-range spin generation over 10$\mu$m, order of magnitudes longer than the electron mean free path $\sim 30$nm, was reported recently \cite{Shiota21}.
Long-range spin profile has been pointed out theoretically to arise in a spin-orbit system with an equal strength of the Rashba and the Dresselhauss interactions \cite{Bernevig06}.
The profile is robust due to a symmetry called a shifting property of the spin-dependent electron dispersion, $\epsilon_-(\kv)=\epsilon_+(\kv+\Qv)$, for a general wave vector $\kv$ and a particular vector $\Qv$, where  $\pm$ is electron spin.
The shifting property leads to a conservation of spin density with a wave vector $\Qv$, namely, spin propagation at $\Qv$ occurs without decay.
It was argued recently that a chiral system of Weyl interaction has the shifting property and long-range transport observed in Ref. \cite{Shiota21} was argued in this context \cite{RoyCM22}.
The infinite decay length is, however, only for the component with the wave vector $\Qv$ and other spin components decay.
The macroscopic spin transport in Ref.  \cite{Shiota21} remains to be a puzzle.

Motivated by the experimental result \cite{Shiota21}, we study theoretically the decay length of spin accumulation in the Weyl system taking account of finite electron elastic lifetime $\tau$ or a finite imaginary part $i\eta$ of energy, $\eta=(2\tau)^{-1}$.
It turns out that the decay length for spin accumulation $\Lambda$ is proportional to $\tau$;
A macroscopic decay length over 10$\mu$m for a mean free path of $30$nm in Ref. \cite{Shiota21} cannot be explained by the present model.

Our study of chiral spin generation is based on a model with the Weyl spin-orbit interaction, described by the field Hamiltonian ($\hat{\ }$ denotes field operator)
\begin{align}
 \hat{H}&=\sumkv \hat{c}_\kv^\dagger \lt(\ekv+\gammav_\kv\cdot\sigmav \rt) \hat{c}_\kv
\end{align}
where $\hat{c}$ and $\hat{c}^\dagger$ are conduction electron field operators with two spin components,  $\sigmav$ is a vector of Pauli matrices, $\ekv\equiv \frac{k^2}{2m}-\ef$ is the free electron energy ($\ef$ is the Fermi energy, $m$ is the electron mass). We use a unit where $\hbar=1$ and $e=1$ ($e$ is the electric charge).
A vector$\gammav_\kv\equiv \lambda \kv$
represents the Weyl spin-orbit interaction with a coupling constant $\lambda$.
The electron velocity operator in the momentum space is
$ \hat{v}_i = \frac{k_i}{m}+ \lambda \sigma_i$,
where the last term is  the anomalous velocity due to the spin-orbit interaction.

The spin density induced by an applied electric field is represented by a correlation function of spin and electric current.
The dominant contribution is \cite{Funaki21}
\begin{align}
 \chi^{(sj)}_{ij}(\qv)&=  \frac{1}{V} \sumkv
 \tr[\sigma_i  G_{\kv}^\ret \hat{v}_j G_{\kv+\qv}^\adv]
 \label{chidef}
\end{align}
where  $V$ is the system volume, $G^\ret_{\kv}=\frac{1}{-\epsilon_k-\gammav_k\cdot\sigmav + i\eta }$ and $ G_{\kv}^\adv = (G_{\kv}^\ret)^* $
are the $2\times2$ matrices of the retarded and advanced Green's function, respectively.
We take account of finite imaginary part $i\eta$ of the energy in the Green's functions arising from elastic impurity scattering.
Evaluating the trace for spin, we obtain
\begin{align}
 \chi^{(sj)}_{ij}(\qv)
 &= \frac{1}{V}\sumkv
 \biggl[
  \lambda \frac{k_j}{m}
  \frac{\lt(k+\frac{q}{2}\rt)_i}{ {\gamma}_{\kv+\frac{\qv}{2}} }
   \Re [f_{\kv-\frac{\qv}{2}}^\ret  h_{\kv+\frac{\qv}{2}}^\adv ]
 +\frac{\lambda}{2}\delta_{ij}  \Re [f_{\kv-\frac{\qv}{2}}^\ret  f_{\kv+\frac{\qv}{2}}^\adv ]
 \nnr
 & +i \lambda^2 \epsilon_{ilm} q_m\frac{k_jk_l}{m}
 \frac{1}{ {\gamma}_{\kv+\frac{\qv}{2}}  {\gamma}_{\kv-\frac{\qv}{2}}}
 \Re [h_{\kv-\frac{\qv}{2}}^\ret  h_{\kv+\frac{\qv}{2}}^\adv ]
 +i \epsilon_{ijl} \lambda^2
  \frac{\lt(k+\frac{q}{2}\rt)_l}{ {\gamma}_{\kv+\frac{\qv}{2}} }
  \Re [f_{\kv-\frac{\qv}{2}}^\ret  h_{\kv+\frac{\qv}{2}}^\adv ]
  \biggr]
 \label{chiexpression}
\end{align}
where $\Re$ denotes the real part,
\begin{align}
f_\kv^\lambda & \equiv \frac{1}{2}\sum_{\sigma=\pm}g_{\kv\sigma}^\lambda , &
h_{\kv}^\lambda &\equiv  \frac{1}{2} \sum_{\sigma=\pm}\sigma g_{\kv\sigma}^\lambda
\end{align}
($\lambda=\ret,\adv$), with
$
g_{\kv\sigma}^\ret \equiv  \frac{1}{-\epsilon_k-\sigma \gamma_k +\frac{i}{2\tau} }
$
being the diagonalized Green's function,
$g_{\kv\sigma}^\adv=(g_{\kv\sigma}^\ret)^* $ ($\sigma=\pm$).

\begin{figure}
\centering
 \includegraphics[width=0.48\hsize]{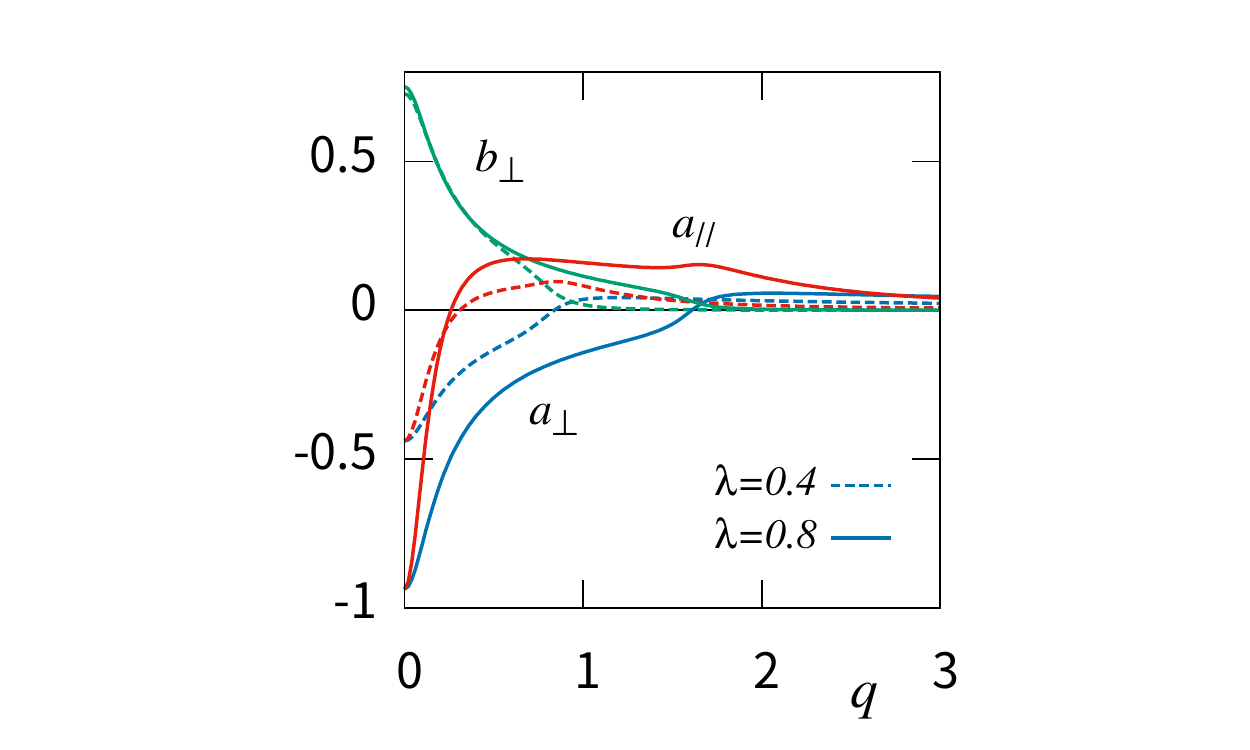}
 \includegraphics[width=0.48\hsize]{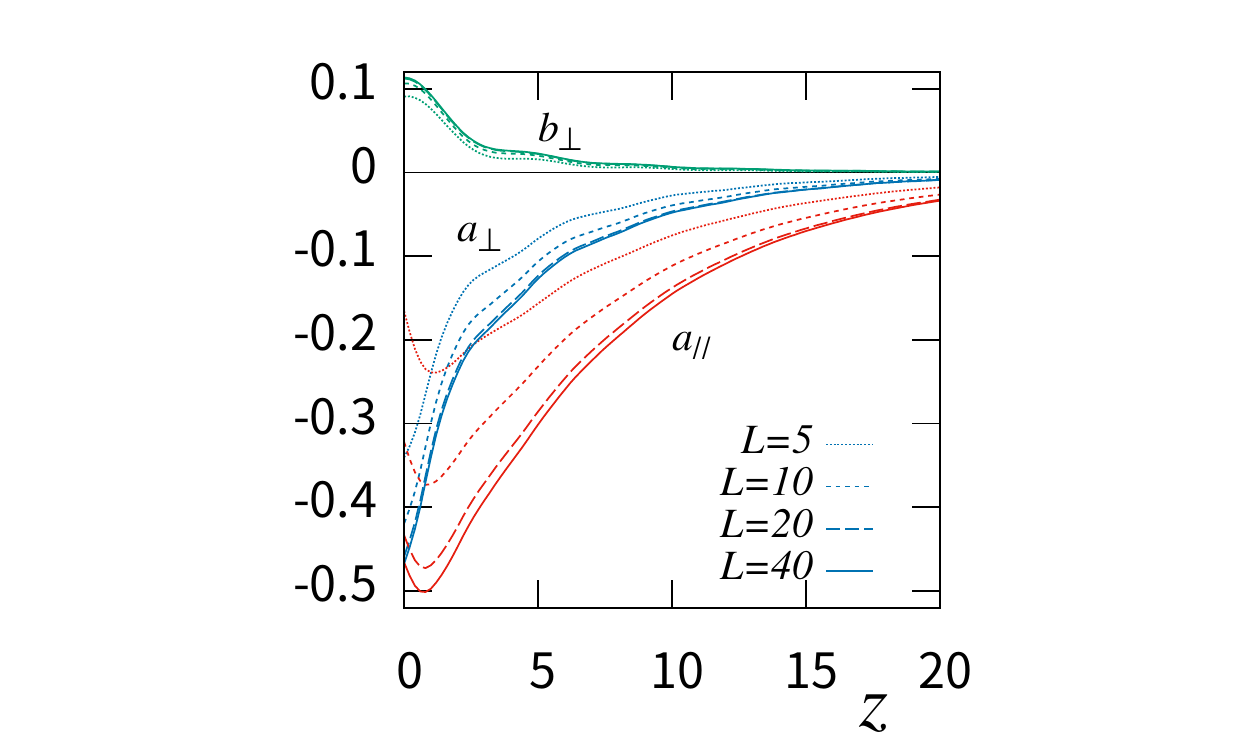}
 \caption{ Left: Coefficients $ a^\parallel(q)$, $ a^\perp(q)$, and $b^\perp(q)$ as functions of wave vector $q/\kf$ for $\lambda/\ef=0.4, 0.8$ at $\eta/\ef=0.1$.
 Right: Spatial profiles of $s_{a\perp}(z)$, $s_{a\parallel}(z)$ and $s_{b\perp}(z)$ (Eq. (\ref{szs})) divided by $E_0$  for a constant applied electric field in the region $-L<z<0$ ($z$ is in unit of $\kf^{-1}$) calculated for $\lambda/\ef=0.8$ and $\eta/\ef=0.1$.
 \label{FIGchiqr}}
\end{figure}

\begin{figure}
\centering
 \includegraphics[width=0.48\hsize]{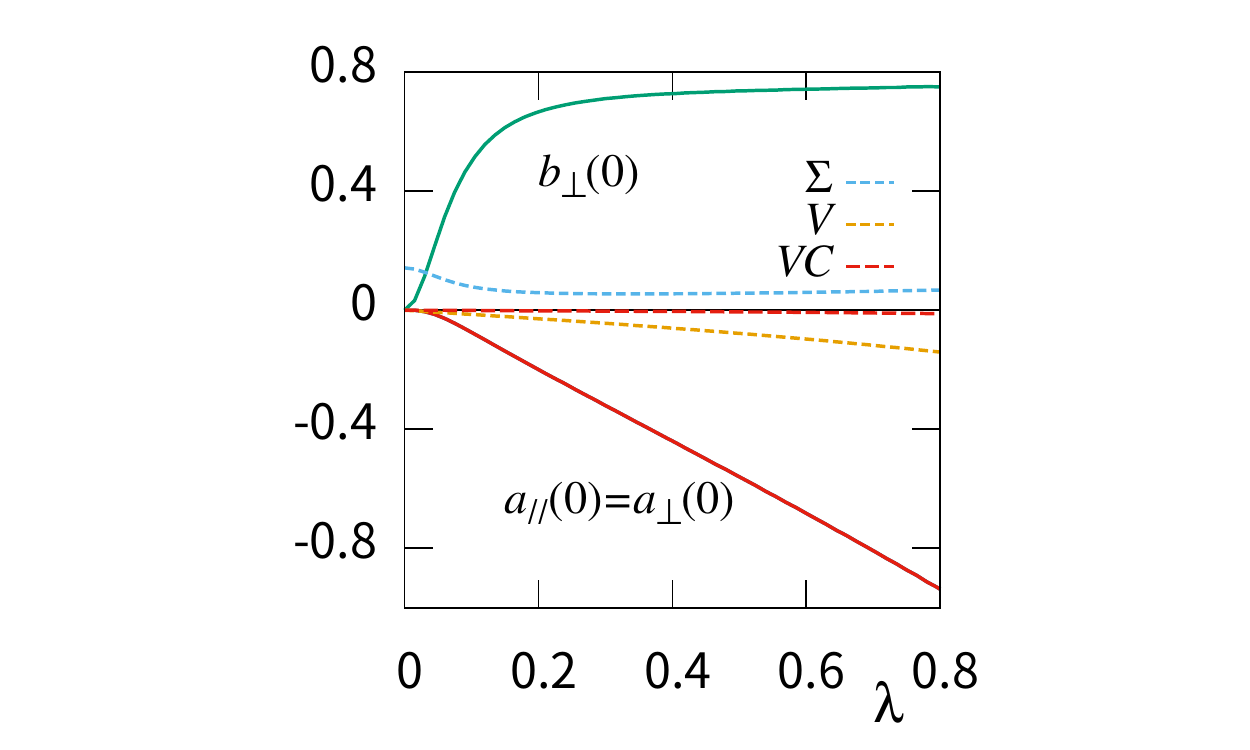}
 \caption{ Uniform ($q=0$)  components of coefficients, $ a^\parallel(0)=a^\perp(0)$ and $b^\perp(0)$  as functions of  spin-orbit coupling $\lambda/\ef$.
 Dashed lines are vertex correction contributions, $\Sigma$, $V$ and the total vertex correction $\chi^{\rm VC}$ (denoted by $VC$).
 \label{FIGchilam}}
\end{figure}

Choosing $\qv$ along the $z$ axis, we obtain
\begin{align}
 \chi^{(sj)}_{ij}(\qv)&= a^\parallel(\qv)\delta_{iz}\delta_{jz}
   +a^\perp(\qv)(\delta_{ij} -\delta_{iz}\delta_{jz} )
   +i\epsilon_{ijk}q_k b^\perp
\end{align}
where
\begin{align}
 a^\parallel(q) & = \frac{\lambda}{V} \sumkv \Re\biggl[
   k_z \frac{k_z+\frac{1}{2}q}{m {\gamma}_{\kv+\frac{\qv}{2}}}
   f_{\kv-\frac{\qv}{2}}^\ret  h_{\kv+\frac{\qv}{2}}^\adv
   + \frac{1}{2}  f_{\kv-\frac{\qv}{2}}^\ret  f_{\kv+\frac{\qv}{2}}^\adv \biggr]
   \nnr
a^\perp(q) & = \frac{\lambda}{V} \sumkv \Re\biggl[
   \frac{k^2}{3m {\gamma}_{\kv+\frac{\qv}{2}}}
   f_{\kv-\frac{\qv}{2}}^\ret  h_{\kv+\frac{\qv}{2}}^\adv
   + \frac{1}{2}  f_{\kv-\frac{\qv}{2}}^\ret  f_{\kv+\frac{\qv}{2}}^\adv \biggr]
   \nnr
b^\perp(q) & = \frac{\lambda^2}{V}\sumkv \Re\biggl[
   \frac{k^2}{3m }
   \frac{1}{ {\gamma}_{\kv+\frac{\qv}{2}}  {\gamma}_{\kv-\frac{\qv}{2}}}
  h_{\kv-\frac{\qv}{2}}^\ret  h_{\kv+\frac{\qv}{2}}^\adv
  + \frac{1}{ 2{\gamma}_{\kv+\frac{\qv}{2}} }
   f_{\kv-\frac{\qv}{2}}^\ret  h_{\kv+\frac{\qv}{2}}^\adv
  \biggr] \label{aabres}
\end{align}
The behaviors  are plotted in Fig. \ref{FIGchiqr}.
Their uniform contributions are
\begin{align}
 a^\parallel(q=0) & = a^\perp (q=0)
 =  \frac{\lambda}{V} \sumkv \Re\biggl[
   \frac{k^2}{3m {\gamma}_{\kv}}
   f_{\kv}^\ret  h_{\kv}^\adv
   + \frac{1}{2} | f_{\kv}^\ret  |^2  \biggr]
   \nnr
b^\perp(q=0) & = \frac{\lambda^2}{V} \sumkv \Re\biggl[
   \frac{k^2}{3m }
   \frac{1}{ {\gamma}_{\kv}^2}
  |h_{\kv}^\ret|^2
  + \frac{1}{ 2{\gamma}_{\kv}}
   f_{\kv}^\ret  h_{\kv}^\adv
  \biggr]
\end{align}
which vanish in the nonchiral case ($\lambda=0$).
Expanding with respect to $\lambda$ using ($g_{\kv}^\ret \equiv g_{\kv\sigma}^\ret |_{\lambda=0}$)
\begin{align}
  f_{\kv}^\ret =&  g_{\kv}^\ret +\gamma_k^2 ( g_{\kv}^\ret)^3+O(\gamma_k^4),
  &  h_{\kv}^\ret =&  \gamma_k (g_{\kv}^\ret)^2+O(\gamma_k^3)
\end{align}
the behaviors near $\lambda=0$ are (neglecting quantities of $o((\ef\tau)^{-1})$)
$ a^\parallel(q=0)  = a^\perp (q=0)
   \simeq -\frac{11}{3}\pi \lambda^3 \nu \kf^2\tau^3
   $ and
   $b^\perp(q=0)
   \simeq  \frac{4}{3}\pi \lambda^2 \nu \kf^2\tau^3
$,
where $\kf$ is the Fermi wave vector, $\nu$ is the electron density of states for $\lambda=0$.
The behaviors of the uniform components as function of chiral parameter $\lambda$  are plotted in Fig. \ref{FIGchilam}.


The expression Eq.(\ref{aabres}) applies to any strength of the Weyl spin-orbit interaction $\lambda$, as the interaction is diagonalized in the spin space and the trace is calculated exactly.
We would like, however,  to avoid arguing the limit of large $\lambda$ (the pure Weyl limit), as the response function  in the continuum diverges at large wave vector if the quadractic (mass) term of the dispersion is neglected, and results depends on the regularization such as introducing the lattice constant.

\begin{figure}
\centering
 \includegraphics[width=0.3\hsize]{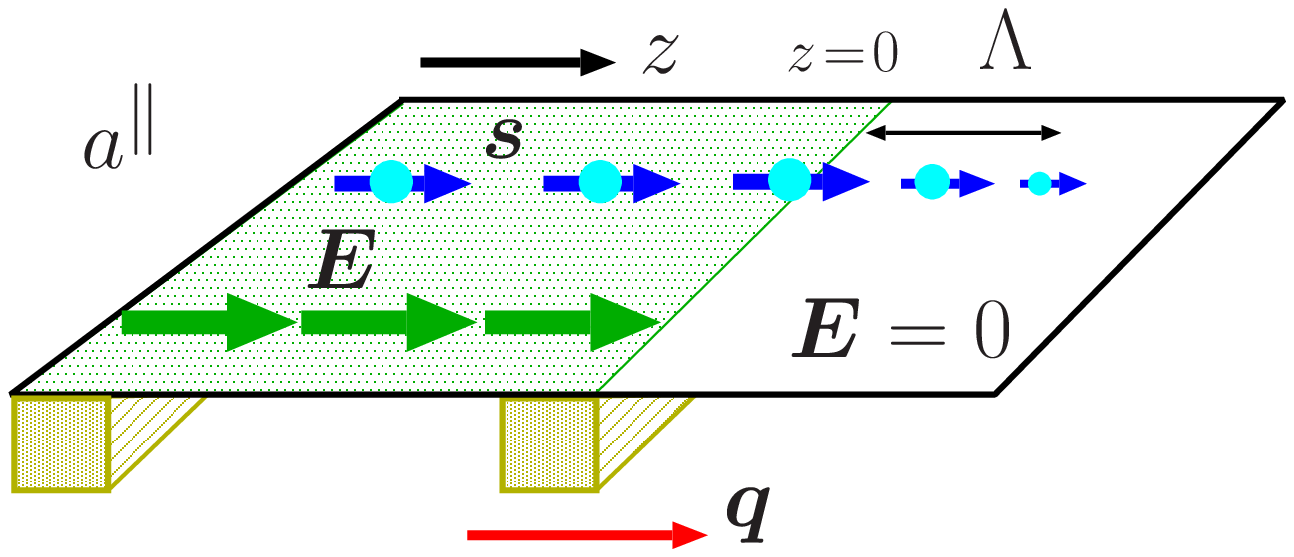}
 \includegraphics[width=0.3\hsize]{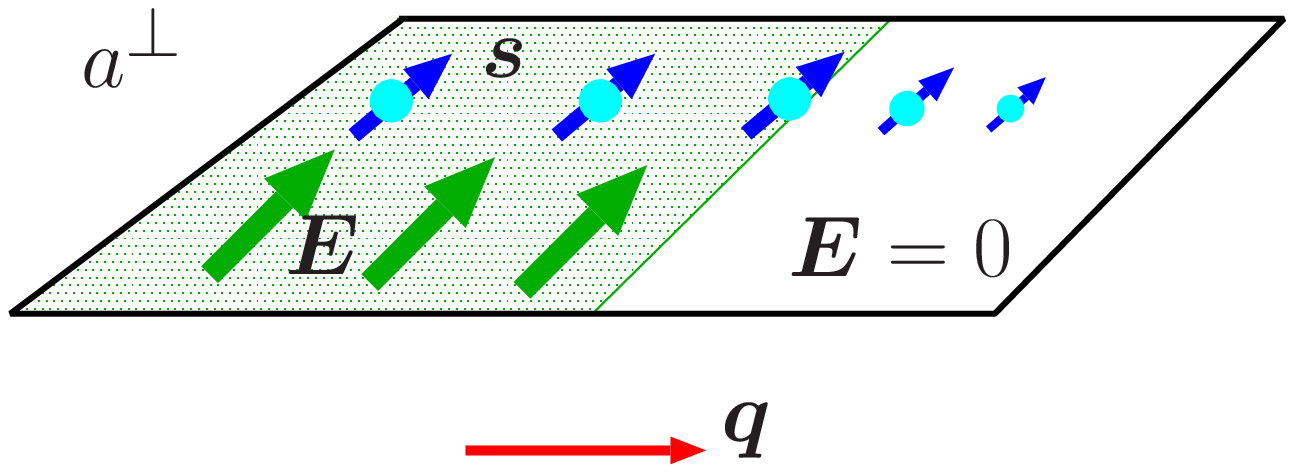}
 \includegraphics[width=0.3\hsize]{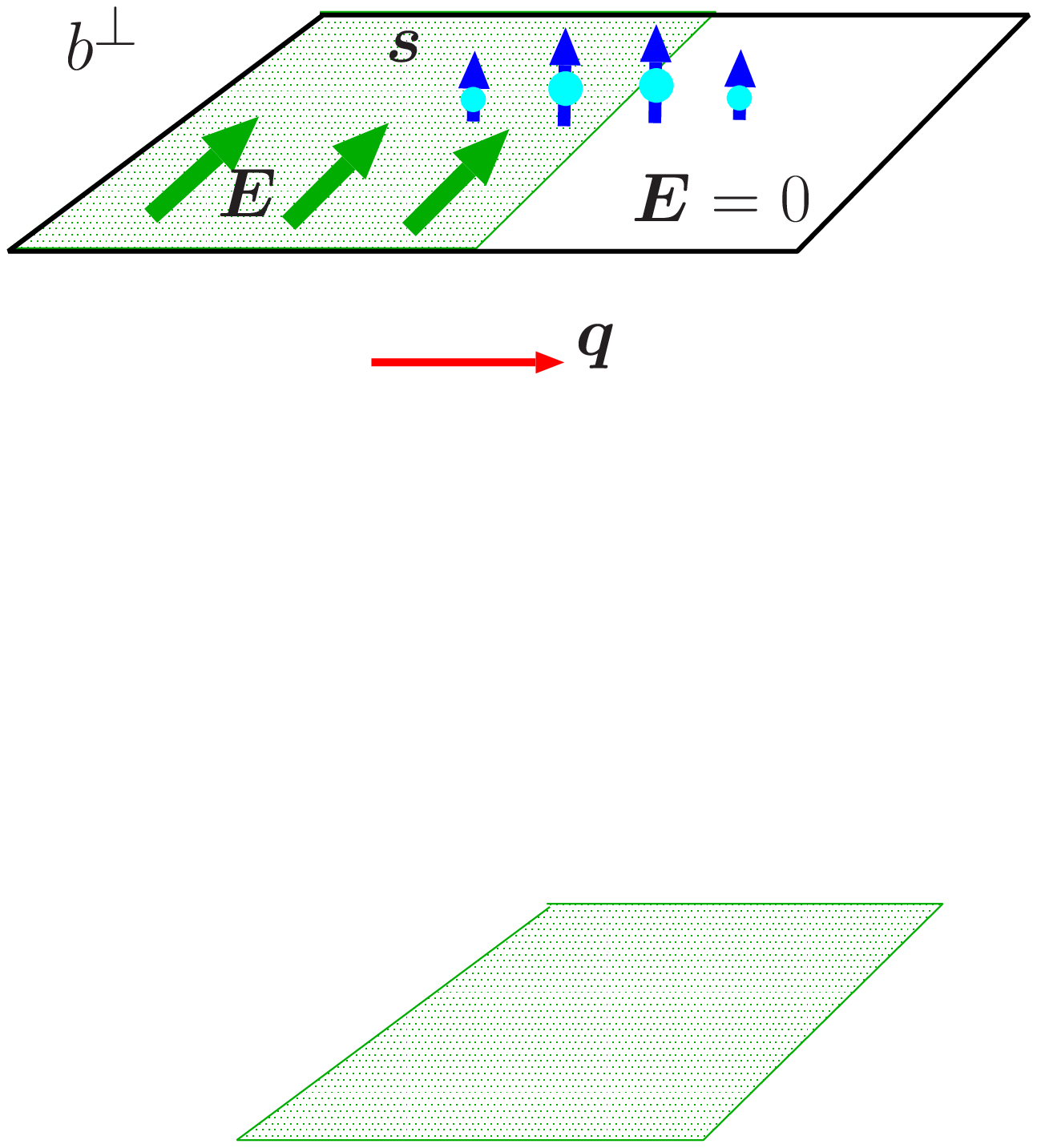}
 \caption{ Schematic figures showing the spin accumulation due to bulk angular momentum generation mechanism represented by $a^\perp$ and $a^\parallel$ and spin Hall mechanism described by $b^\perp$.
 Arrows with $\qv$ denote the direction of variation of the applied electric field $\Ev$ near $z=0$.
 Configuration for $a^\parallel$ is the one used in experiment of Ref. \cite{Shiota21} by attaching two leads (shown below the system plane) at $z=0$ and $z<0$.
 \label{FIGsummary}}
\end{figure}

In the real space representation with an applied electric field changing along the $z$ axis, the induced spin accumulation profile,
$s_i(z)= \int\frac{dq}{2\pi} e^{iqz}  \chi^{(sj)}_{ij}(q)E_j(q)$, is
$\sv=\sv_{a\perp}+\sv_{a\parallel}+\sv_{a\perp}$, where
\begin{align}
 \sv_{a\perp}(z) &= \int\frac{dq}{2\pi} e^{iqz}
   a^\perp(q)(\Ev(q)-\zvhat(\zvhat\cdot\Ev(q))  \nnr
 \sv_{a\parallel}(z) &= \int\frac{dq}{2\pi} e^{iqz}
   a^\parallel(q) \zvhat(\zvhat\cdot\Ev(q))  \nnr
 \sv_{b\perp}(z) &= \int\frac{dq}{2\pi} e^{iqz}
   b^\perp(q)(-i)(\qv\times\Ev(q))
   \label{szs}
\end{align}
represent the induced spin along ($\sv_{a\perp}$ and $\sv_{a\parallel}$) and  perpendicular ($\sv_{b\perp}$) to $\Ev$, respectively.
The contributions $a^\perp$ and $a^\parallel$  represent the bulk angular momentum ($\Lv$) generation effect \cite{Funaki21}  of chiral system, $\Lv\propto \Ev$, and
$b^\perp$ represents the spin Hall effect, where spin emerges proportional to $\nabla\times\Ev$ in the long-wavelength limit \cite{TataraSH18}.
The configurations for $a^\perp$, $a^\parallel$ and $b^\perp$ are shown in Fig. \ref{FIGsummary}.

\begin{figure}
\centering
 \includegraphics[width=0.48\hsize]{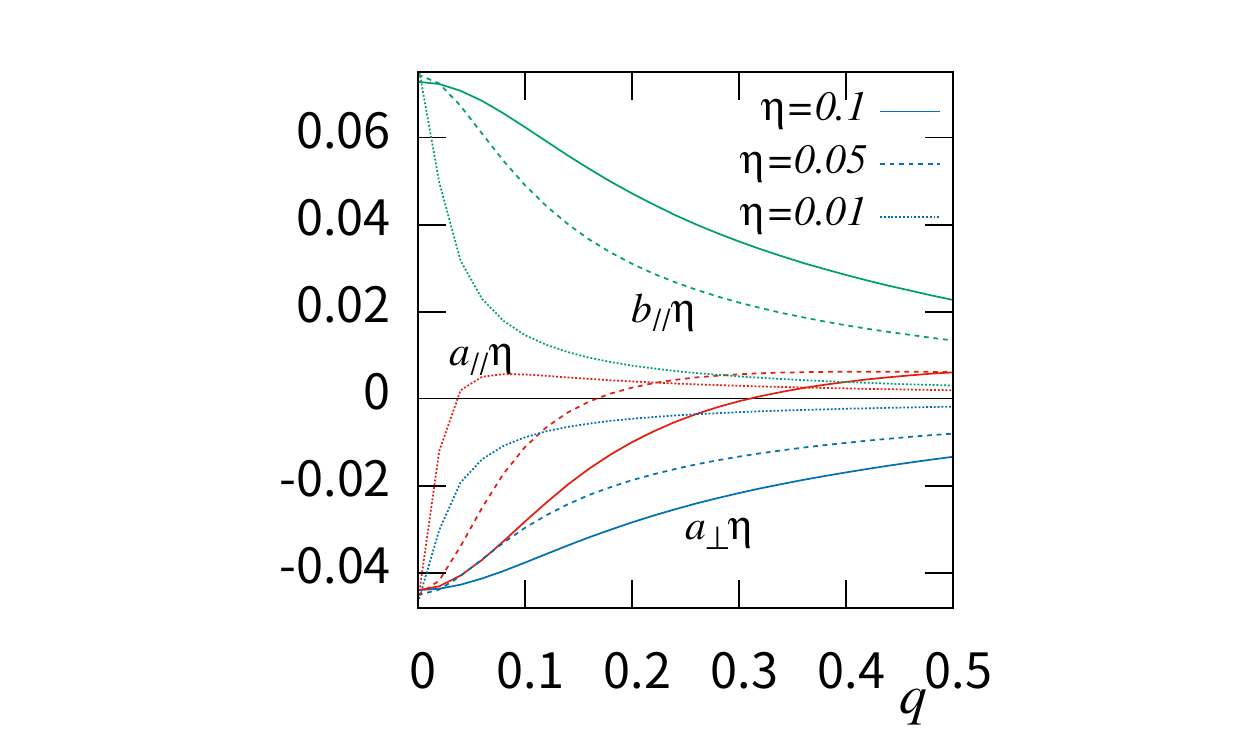}
 \caption{ Dependence of coefficients, $ \eta a^\parallel(q)$, $\eta a^\perp(q)$ and $\eta b^\perp(q)$ on $\eta$ at $\lambda/\ef=0.4$.
 \label{FIGchiq_eta}}
\end{figure}

\begin{figure}
\centering
 \includegraphics[width=0.48\hsize]{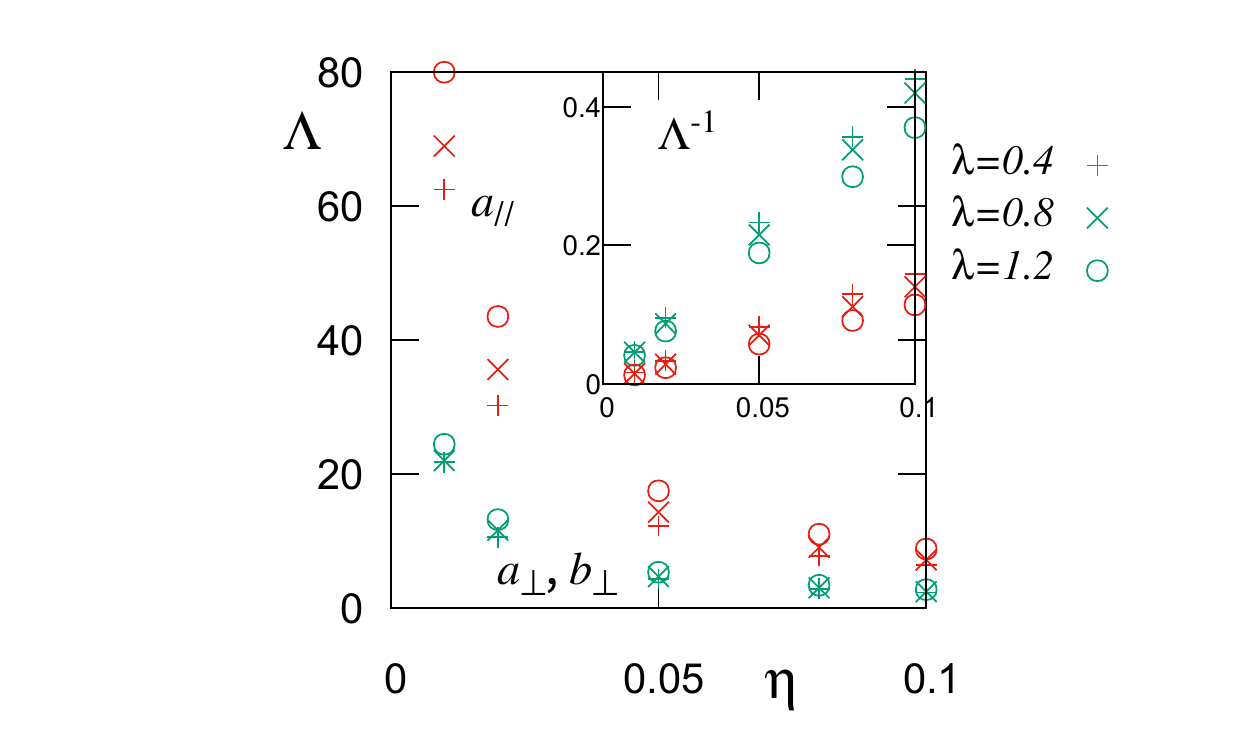}
 \caption{ Decay length $\Lambda$ for $ a^\parallel$, $a^\perp$ and $b^\perp$ determined by the width near $q=0$ plotted as function of $\eta$.
 $\Lambda_{a\perp}=\Lambda_{b\perp}$.
 Inset: Plot of $\Lambda^{-1}$ showing linear dependence on $\eta$.
 \label{FIGLam}}
\end{figure}

\begin{figure}
\centering
 \includegraphics[width=0.48\hsize]{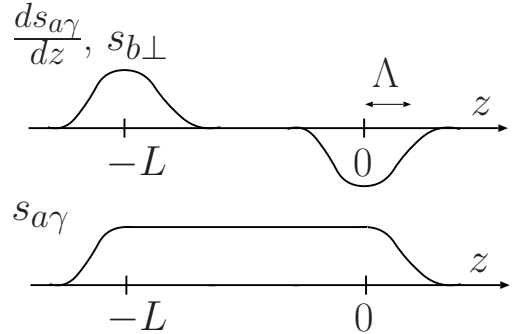}
 \caption{ Schematic figure showing the spin accumulation profile  $s_{a\gamma}(z)$ ($\gamma=\parallel,\perp$),   $s_{b\perp}(z)$ and derivative $\frac{ds_{a\gamma}}{dz}$ (Eq. (\ref{saabresultapprox}))  for $\Lambda\ll L$.
 \label{FIGspinaccum}}
\end{figure}

Generation of spin density by an applied electric field requires a mechanism to convert a linear momentum to an angular momentum. In the present model, the Weyl spin-orbit interaction plays this role, and it is natural that both orbital angular momentum and spin have non-vanishing uniform response to an electric field in chiral systems. One a uniform ($q=0$) response survives, nonlocal responses are generally expected (See Eq. (\ref{saabresultapprox})).
To explore the nonlocal nature of the induced spin clearly, we consider the case of an applied electric field in a restricted region, $-L<z<0$, i.e., $\Ev=\Ev_0(\theta(-z)-\theta(-z-L))$, where $\Ev_0$ is a constant vector and $\theta(x)$ is the step function.
The case of $\Ev_0\parallel \zvhat$ corresponds to a configuration in the experiment of Ref. \cite{Shiota21}, with two leads attached at $z=-L$ and $z=0$ and an applied voltage between them.
(The electric field near the lead is continuous towards the thickness direction and no electric charge appears in ideally smooth structures.)
The Fourier transform of the applied electric field is $\Ev(q)=\frac{\Ev_0}{-iq}(1-e^{iqL})$, and the amplitudes of the spin accumulation read ($\gamma=\parallel,\perp$)
\begin{align}
 s_{a\gamma}(z)
 & = \frac{E_0}{2\pi} \int_{-\infty}^\infty dq \frac{ie^{iqz}}{q}(1-e^{iqL})  a^\gamma(q)
\nnr
%
   %
s_{b\perp}(z)
 & = \frac{E_0}{2\pi} \int_{-\infty}^\infty dq e^{iqz} (1-e^{iqL})  b^\perp(q)
 \label{saabresult}
\end{align}
As $a^\perp(q)$, $a^\parallel(q)$ and $ b^\perp(q)$ are peaked near $q=0$ (Fig. \ref{FIGchiqr}), we approximate the long-range profile by the Gaussians as
$a^\gamma(q)\simeq a^\gamma(0)e^{-(q\Lambda_{a\gamma})^2}$,
and $ b^\perp(q)\simeq b^\perp(0)e^{-(q\Lambda_{b\perp})^2}$ with width $ \Lambda_{a\gamma}^{-1}$
and $ \Lambda_{b\perp}^{-1}$, respectively.
As seen in Figs. \ref{FIGchiq_eta} and \ref{FIGLam}, the magnitudes at $q=0$ and width scale as $a^\gamma(0), b^\perp(0)\propto \eta^{-1}$ and $ \Lambda_{a\gamma}, \Lambda_{b\perp}\propto \eta $.
Decay length $\Lambda$ for the three components are plotted as function of $\eta$ in Fig. \ref{FIGLam}.
Within this approximation, we obtain
\begin{align}
 \frac{ds_{a\gamma}}{dz}
 & \simeq
 -\frac{E_0a^\gamma(0)}{2\Lambda_{a\gamma}}
 \lt[e^{-\frac{z^2}{4\Lambda_{a\gamma}^2}}-e^{-\frac{(z+L)^2}{4\Lambda_{a\gamma}^2}}\rt]
\nnr
%
   %
s_{b\perp}(z)
 & \simeq \frac{E_0b^\perp(0)}{2\Lambda_{b\perp}}
 \lt[e^{-\frac{z^2}{4\Lambda_{b\perp}^2}}-e^{-\frac{(z+L)^2}{4\Lambda_{b\perp}^2}}\rt]
 \label{saabresultapprox}
\end{align}
The spin profiles of $s_{a\gamma}(z)$  are as in the lower figure of Fig. \ref{FIGspinaccum}, decaying with the corresponding decay length away from the region of finite applied field.
The spin Hall component  $s_{b\perp}(z)$ in contrast is localized near the edge of the region of the applied field.

Besides the contribution considered above, there is a contribution called the vertex correction, whose lowest order contribution is depicted in Fig. \ref{FIGchilam}.
The contribution to the response function at $q=0$ reads
\begin{align}
 \chi^{\rm VC}_{ij} &= n_{\rm i}v_{\rm i}^2\tr[\Sigma_i V_j]
\end{align}
where $n_{\rm i}$ and $v_{\rm i}$ are the concentration and potential strength of impurities ($\pi\nu n_{\rm i}v_{\rm i}^2=\eta$) and
\begin{align}
\Sigma_i &\equiv \frac{1}{V}\sum_{\kv} G_{\kv}^\adv \sigma_i G_{\kv}^\ret ,&
V_j &\equiv \frac{1}{V}\sum_{\kv} G_{\kv}^\adv v_j G_{\kv}^\ret
\end{align}
are the correction matrices to the spin and velocity vertices, respectively.
Using the rotational symmetry with respect to $\kv$, they read $\Sigma_i =\Sigma \sigma_i$ and $V_j=V\sigma_j$, where
\begin{align}
\Sigma &=\frac{1}{V} \sum_{\kv}\lt[ |f_{\kv}^\adv|^2 -\frac{1}{3} |h_{\kv}^\adv|^2 \rt]
   \nnr
V &= -\frac{2\lambda}{3m} \frac{1}{V}\sum_{\kv}\frac{k^2\epsilon_k}{\epsilon_k^2+\eta^2 } |f_{\kv}^\adv|^2
\end{align}
The correction at $q=0$ is therefore  $\chi^{\rm VC}_{ij} =\chi^{\rm VC} \delta_{ij}$ with
$\chi^{\rm VC}=\frac{3\eta}{\pi}\Sigma V$.
The vertex correction turns out to be negligibly small compared to the contribution $a^\parallel$ as shown in Fig. \ref{FIGchilam}.

To summarize, we have calculated the response function of spin in a Weyl electron system for an applied electric field at finite external wave vector $q$.
It was shown that spin-current response of the chiral system has a finite uniform ($q=0$) component as has been mentioned previously \cite{Funaki21}.
The $q$-space structure of the response function indicates that electrically-induced spin accumulation in chiral systems show nonlocal nature, i.e., the nonequilibrium spin density emerges away from the region where the electric field or current is applied.
The decay length of spin accumulation, which corresponds to the spin diffusion length, is proportional to the electron elastic mean free path.
Macroscopically nonlocal  spin polarization recently observed in chiral conductors NbSi$_2$ and TaSi$_2$ with short mean free paths is not explained by the present model.

\acknowledgements
The author thank J. Kishine and Y. Togawa  for valuable discussion
and  Foundation Advanced Technology Institute (ATI) for a support which initiated this work.
This study was supported by
a Grant-in-Aid for Scientific Research (B) (No. 21H01034) from the Japan Society for the Promotion of Science.



\end{document}